\documentclass[aps,prb,reprint,amsmath]{revtex4-2}
\usepackage{hyperref}
\usepackage{color}
\usepackage[caption=false]{subfig}

\usepackage{graphicx,epstopdf}
\usepackage{bm}
\usepackage{lipsum}

\begin{document}
\title{{\Large{}{}Nonthermal Dynamics and Scar-Like Spectral Structures in a High-Spin Fermi Gas}}

\author{Shuyi Li}

\affiliation{Institute of Theoretical Physics, University of Science and Technology Beijing, Beijing 100083, China}
\author{Qiang Gu}
\email[Corresponding author: ] {qgu@ustb.edu.cn}
\affiliation{Institute of Theoretical Physics, University of Science and Technology Beijing, Beijing 100083, China}

\begin{abstract}

We investigate nonequilibrium dynamics and weak ergodicity breaking in a harmonically trapped spin-$3/2$ Fermi gas by using the time-dependent Hartree–Fock equation. The Shannon entropy remains bounded and oscillatory throughout the evolution, indicating restricted and nonuniform exploration of Hilbert space rather than immediate thermalization. The fidelity exhibits pronounced, nearly periodic revivals whose period is largely insensitive to particle number and interaction strength, while the revival amplitude gradually decreases with increasing system size and interaction strength. The Fourier spectrum of the fidelity reveals a set of sharp and approximately equally spaced peaks. By projecting the time-evolved state onto the instantaneous eigenbasis of the self-consistent mean-field Hamiltonian, we identify a sparse and spectrally stable manifold that forms a quasi-regular energy ladder, with spacing comparable to the dominant quasienergy interval extracted from the fidelity spectrum. These results indicate that the long-lived coherent oscillations originate from collective phase interference associated with a quasi-regular spectral structure embedded in the many-body continuum, rather than from a conventional eigenstate-dominated scar mechanism.

\end{abstract}
\maketitle

\section{Introduction}

Understanding how isolated quantum many-body systems approach—or fail to approach—thermal equilibrium remains a central question in nonequilibrium quantum physics. The eigenstate thermalization hypothesis (ETH) posits that, in generic nonintegrable systems, the expectation values of local observables in individual eigenstates coincide with the corresponding thermal ensemble averages\cite{Deutsch-1991,Srednicki-1994,Rigol-2008}. However, in recent years, remarkable deviations from ETH have been identified, revealing distinct mechanisms of ergodicity breaking. Two paradigmatic examples are many-body localization (MBL), where quenched disorder prevents thermalization, and quantum many-body scars (QMBS), where a small subset of nonthermal eigenstates embedded in an otherwise thermal spectrum can give rise to long-lived nonergodic dynamics for certain initial states \cite{Abanin-2019,Serbyn-2021}.

The notion of ``scars'' originates from single-particle quantum chaos, where certain eigenfunctions exhibit enhanced probability density along unstable classical periodic orbits \cite{Heller-1984}. Extensions of this idea to interacting systems have revealed the existence of atypical eigenstates that weakly violate ETH while coexisting with thermal states. Approximate many-body scars were first observed experimentally in Rydberg-atom quantum simulators \cite{Bernien-2017}, effectively described by the PXP model, which exhibits coherent revivals from specially prepared initial states \cite{Turner-2018}. Since then, scar-like dynamical behavior has been reported in a range of lattice models and quantum simulators, including Bose–Hubbard and Su–Schrieffer–Heeger–type systems \cite{Su-2023,Zhang-2023}. Theoretical studies have further explored the structure, stability, and dynamical consequences of such weakly nonergodic states  \cite{Turner-2018,Schecter-2019,Ho-2019,Lin-2019,Lin-2020,Michailidis-2020,Moudgalya-2020,Kuno-2021,Surace-2021,Ren-2022,Lerose-2025,Mukherjee-2020,Pakrouski-2021,Desaules-2023,Mohapatra-2023,Wang-2024,HuangW-2025,Huang-2025,Moudgalya-2022,Chen-2024,Choi-2019,Alhambra-2020,Iadecola-2020,Chioquetta-2025}.

Despite these advances, most existing studies have focused on lattice-based models, whereas continuum many-body systems remain comparatively less explored in the context of quantum scarring. High-spin Fermi gases constitute a particularly intriguing setting due to their enlarged spin degrees of freedom and intrinsic spin-mixing interactions. Experiments and theoretical analyses have reported giant, slowly damping collective oscillations in such systems \cite{FengT-2017,Dong-2013,Lisy-2025,Krauser-2014}, suggesting that coherent nonthermal behavior can persist over extended timescales. This raises a natural question: whether such long-lived oscillations can be associated with a sparse, weakly hybridized spectral structure analogous to scar manifolds identified in lattice systems?

In our previous study \cite{Lisy-2025}, we reported robust nonthermal oscillations in trapped high-spin Fermi gases that cannot be readily explained within the conventional eigenstate thermalization hypothesis. Here we investigate their microscopic origin using the time-dependent Hartree–Fock (TDHF) equation for a harmonically trapped spin-$3/2$ Fermi gas. Starting from an imbalanced spin configuration, the Shannon entropy exhibits long-lived oscillations, indicating delayed and nonuniform exploration of the Hilbert space rather than immediate equilibration. The fidelity displays pronounced revivals characterized by a stable dominant frequency that persists throughout the evolution, and its Fourier spectrum exhibits sharp peaks reflecting a stable interference pattern set by quasi-regular energy spacings. Analysis of the instantaneous self-consistent spectrum further uncovers a subset of approximately equispaced modes embedded within the dense many-body background. The characteristic spacing of this quasi-regular spectral subset is comparable to the dominant frequency extracted from the fidelity dynamics. These results suggest that the observed long-lived oscillations originate from a quasi-regular, scar-like spectral structure embedded in the many-body continuum. While sharing certain phenomenological similarities with quantum many-body scars, the present behavior arises from collective interference among a weakly hybridized set of modes in a finite, self-consistent continuum system rather than from a small set of dominant eigenstates.

\section{Model and method}

We consider a harmonically trapped spin $f=3/2$ Fermi gas, which provides a minimal continuous platform for studying nonthermal dynamics. 

The Hamiltonian of the system is ${\bm H}={\bm H_0}+{\bm H_I}$:
\begin{eqnarray}
	{\bm H}&=&-\frac{1}{2}\nabla^{2}+\frac{1}{2}{x}^{2}\notag\\
	&=&
	+\sum_{F=0}^{2f-1} \sum_{m_F=-F}^F g_F\delta(x-x')|F,m_F\rangle\langle F,m_F|.
\end{eqnarray}The first term on the right side describes the kinetic energy, the second term is the potential energy with $x=\tilde{x}/\sqrt{\hbar/(m\omega)}$, $\tilde{x}$ is the position of the particle and $\omega$ is the frequency of the $1D$ harmonic trap. The last term is the two-body interaction $H_I$ with the reduced two-body interaction strength $g_{F}={\tilde{g_F}}/({\hbar\omega \sqrt{\hbar/m\omega}})$, which is connected to the $s$-wave scattering length $a_F$, a positive $a_F$ can be regarded as a repulsive interaction in the low-energy scattering states. $F$ represents the total spin channel and $F=0,1,2,3$, since we only consider $s$-wave interaction, for Fermi gases, the space wavefunction is symmetric, and the spin wavefunction must be antisymmetric. Therefore, only the even total spin $(F=0,2)$ channels exist. This means that only two interacting channel parameters, $g_0$ and $g_2$, are considered. The magnetic number $m_F=m_1+m_2$ is conserved and the total spin $F$ can vary between $0$ and $2$.

The present model is not known to be integrable. Although certain one-dimensional fermionic models are exactly solved in translationally invariant settings, the harmonic confinement considered here breaks translational invariance, thereby precluding Bethe-ansatz integrability\cite{Rigol-2008}. Moreover, for spin-$3/2$ fermions, integrability or enhanced symmetry arises only at special fine-tuned interaction points, whereas the generic interaction parameters used in this work lie away from such limits\cite{Wu-2003}.

Each energy level is fourfold degenerate due to spin components with $m = \pm1/2$ and $m = \pm3/2$. The many-body wave function of the system $\Psi(\bm x)$ can be written as a Slater determinant,

\begin{align}
	\Psi(\bm x)&=&\renewcommand{\arraystretch}{1.3}\setlength{\arraycolsep}{8pt}
	\frac{1}{\sqrt{(2n)!}} \left|\begin{matrix}
		{\psi}_{1}^{\uparrow}(x_{1})&{\psi}_{1}^{\uparrow}(x_{2})&\cdots& {\psi}_{1}^{\uparrow}(x_{2n})\\
		{\psi}_{1}^{\downarrow}(x_{1})&{\psi}_{1}^{\downarrow}(x_{2})&\cdots& {\psi}_{1}^{\downarrow}(x_{2n})\\
		\vdots&\vdots&\ddots&\vdots\\
		{\psi}_{n}^{\uparrow}(x_{1})&{\psi}_{n}^{\uparrow}(x_{2})&\cdots& {\psi}_{n}^{\uparrow}(x_{2n})\\
		{\psi}_{n}^{\downarrow}(x_{1})&{\psi}_{n}^{\downarrow}(x_{2})&\cdots& {\psi}_{n}^{\downarrow}(x_{2n})
		\end{matrix} \right|,
\end{align}
$\bm{x}=(x_1,x_2,\cdots,x_{2n})$, there are four degenerate single-particle states for each energy level $n$, but we consider only two of them
$(s=\uparrow,\downarrow)$.

\begin{eqnarray}
	{\psi}_{n}^{s}(x) &=& \sum_{m=\pm3/2,\pm1/2} {\alpha}_{n,m}^{s}\varphi_n(x)\chi_m \notag\\
	&=&
	 \sum_{m=\pm3/2,\pm1/2} \phi_{n,m}^{s}(x)\chi_m,
	\label{psi}
\end{eqnarray}
where ${\phi}_{n,m}^{s}(x)=\alpha_{n,m}^s\varphi_n(x)$, $\alpha_{n,m}^s$ is the superposition coefficient of the four pure states, $\varphi_n(x)$ is the eigenfunction of the harmonic oscillator and $\chi_m$ is the spin basis.

The energy of the system is
\begin{eqnarray}
	E&=&\int \Psi^{*}(\bm x)(H_0+H_I)\Psi(\bm x)d\bm x=E_0+E_I.
\end{eqnarray}
 The TDHF equations are obtained by varying $(E - i\hbar\partial_t N)$ to $\psi_{n}^{s}$, 
  
\begin{eqnarray}
	(i\hbar\partial_t - H_0)\psi_{n}^{s}(x) = \sum_{n^{\prime}, s^{\prime}} {\bm M}_{n^{\prime}}^{s^{\prime}}\psi_{n}^{s}(x),
\end{eqnarray}
where $H_0$ denotes the single-particle Hamiltonian, and ${\bm M}_{n^{\prime}}^{s^{\prime}}$ accounts for the mean-field interaction effects, including the spin-mixing term characterized by $c_2$ and the non-spin-mixing term $c_0$. Essentially, they are not independent because $c_2 \sim g_2-g_0$ and $c_0 \sim g_2+g_0$. Moreover, all the variables are set to be dimensionless for calculation convenience.

\begin{widetext}
	\begin{eqnarray}
		\centering
		{\bm M}_{n^\prime}^{s^\prime}&=&\renewcommand{\arraystretch}{1.5}\setlength{\arraycolsep}{-0.8pt}
		\left(\begin{matrix}
			c_0|\phi_{n^{\prime},-3/2}^{s^{\prime}}|^{2}                                    &
			(c_2-c_0)\phi_{n^{\prime},-3/2}^{s^{\prime}\ast} \phi_{n^{\prime},-1/2}^{s^{\prime}}&
			-(c_2+c_0)\phi_{n^{\prime},-3/2}^{s^{\prime}\ast} \phi_{n^{\prime},1/2}^{s^{\prime}}&
			-c_0\phi_{n^{\prime},-3/2}^{s^{\prime}\ast} \phi_{n^{\prime},3/2}^{s^{\prime}} \\
			(c_2-c_0)\phi_{n^{\prime},-1/2}^{s^{\prime}\ast} \phi_{n^{\prime},-3/2}^{s^{\prime}}  &
			c_0|\phi_{n^{\prime},-1/2}^{s^{\prime}}|^{2}                                      &
			-c_0\phi_{n^{\prime},-1/2}^{s^{\prime}\ast} \phi_{n^{\prime},1/2}^{s^{\prime}}    &
			-(c_2+c_0)\phi_{n^{\prime},-1/2}^{s^{\prime}\ast} \phi_{n^{\prime},3/2}^{s^{\prime}} \\
			-(c_2+c_0)\phi_{n^{\prime},1/2}^{s^{\prime}\ast} \phi_{n^{\prime},-3/2}^{s^{\prime}}  &
			-c_0\phi_{n^{\prime},1/2}^{s^{\prime}\ast} \phi_{n^{\prime},-1/2}^{s^{\prime}}    &
			c_0|\phi_{n^{\prime},1/2}^{s^{\prime}}|^{2}                                    &
			(c_2-c_0)\phi_{n^{\prime},1/2}^{s^{\prime}\ast} \phi_{n^{\prime},3/2}^{s^{\prime}} \\
			-c_0\phi_{n^{\prime},3/2}^{s^{\prime}\ast} \phi_{n^{\prime},-3/2}^{s^{\prime}}      &
			-(c_2+c_0)\phi_{n^{\prime},3/2}^{s^{\prime}\ast} \phi_{n^{\prime},-1/2}^{s^{\prime}}&
			(c_2-c_0)\phi_{n^{\prime},3/2}^{s^{\prime}\ast} \phi_{n^{\prime},1/2}^{s^{\prime}} &
			c_0|\phi_{n^{\prime},3/2}^{s^{\prime}}|^{2}
		\end{matrix}\right).
		\label{M}
	\end{eqnarray}
\end{widetext}

The system is initialized in a highly excited configuration, with each energy level doubly occupied by fermions predominantly in the $m=\pm1/2$ spin states. The corresponding single-particle wave functions are expressed as
\begin{eqnarray}
	\psi_{n}^{\uparrow}(x) &=& e^{-iS_x\theta}[0,1,0,0]^{T}\varphi_n(x), \notag\\
	\psi_{n}^{\downarrow}(x) &=& e^{-iS_x\theta}[0,0,1,0]^{T}\varphi_n(x),
\end{eqnarray}
where $S_x$ is the spin-$3/2$ matrix and $\theta=0.05$ represents a small rotation angle that controls the initial spin superposition. This preparation introduces a weak mixing among the spin-$z$ eigenstates while keeping the population concentrated in the $m = \pm1/2$ manifold. In such initial state, the relative phases among spin components and orbitals are fixed according to the above construction. The many-body Slater determinant constructed from these single-particle wave functions defines the initial state $\psi(0)$. With time evolution, the spin-mixing interactions ($c_2 \neq0$) coherently couple the $m = \pm1/2$ and $m = \pm3/2$ components, driving oscillatory population transfer between them. The resulting dynamics display long-lived collective oscillations that are characteristic of nonthermal evolution.

\section{Results and discussion}

We analyze the nonequilibrium spin dynamics of the trapped spin-$3/2$ Fermi gas by monitoring both the population evolution of each spin component and the associated Shannon entropy. This approach allows us to connect the macroscopic spin-mixing oscillations with the degree of redistribution among spin components.

The Shannon entropy of the spin populations is defined as
\begin{equation}
	S(t) = -\sum_{m} n_m(t)\,\log n_m(t),
\end{equation}
where $n_m(t)$ denotes the normalized occupation of spin component $m=\pm1/2, \pm3/2$, satisfying $\sum_m n_m(t)=1$. In a fully thermalized configuration with equal spin populations, the entropy reaches its maximal value $S_{\mathrm{th}}=\log 4$.

\begin{figure}
	\begin{center}
		\includegraphics[width=0.9\linewidth]{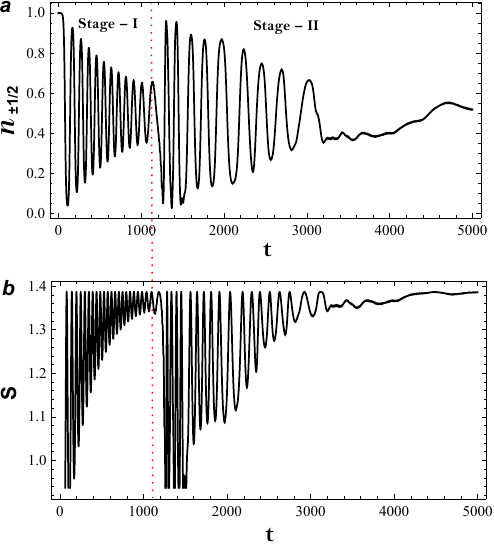}
	\end{center}
	\caption{\label{entropy} {\bf Nonthermal spin-mixing dynamics and Shannon entropy.} Time evolution of relative populations $n_m(t)$ for $m = \pm1/2$ ({\bf a}), where $\Delta T$ is the oscillation period, and the corresponding Shannon entropy $S(t)$ ({\bf b}). The particle number is $N=40$, with interactions $c_0=1.0$ and $c_2=0.1$.}
\end{figure}

Fig.\ref{entropy} shows the time evolution of the relative populations and the corresponding Shannon entropy. At early times (Stage I), the populations exhibit regular, well-defined oscillations, accompanied by synchronized oscillations of $S(t)$. At later times (Stage II), the oscillations become less regular and the entropy develops a slowly increasing envelope. The minima of $S(t)$ gradually rise, reflecting progressive redistribution among spin components. For sufficiently long evolution times, $S(t)$ approaches the thermal limit $\log 4$, signaling a gradual approach toward thermalization rather than permanent localization in a restricted subspace.

We note that the bounded entropy observed at intermediate times does not imply Hilbert space fragmentation\cite{Sala-2020,Khemani-2020}. In genuine fragmentation scenarios, the Hilbert space decomposes into dynamically disconnected sectors, leading to persistent confinement and strictly limited entropy growth. In contrast, the gradual upward drift of the entropy minima and the eventual saturation toward $\log 4$ indicate progressive hybridization with surrounding modes, rather than structural isolation within a disconnected subspace.

While the entropy dynamics demonstrate delayed thermalization and long-lived oscillatory behavior, this coarse-grained quantity alone does not reveal the spectral origin of the coherence. To further probe the dynamical structure underlying the revivals, we next analyze the fidelity,
\begin{equation}
	P(t) = |\langle \psi(0) | \psi(t) \rangle|^2,
\end{equation}
quantifies the return probability of the evolving state to the initial state. In systems exhibiting weak ergodicity breaking or scar-like dynamical structures, fidelity revivals provide a sensitive probe of long-lived coherence.\cite{Turner-2018,Choi-2019,Alhambra-2020,Serbyn-2021,Chioquetta-2025} Pronounced and recurrent peaks in $P(t)$ indicate that the accumulated dynamical phases repeatedly align to produce partial reconstruction of the initial state, rather than undergoing rapid and irreversible dephasing.

\begin{figure}
	\begin{center}
		\includegraphics[width=0.9\linewidth]{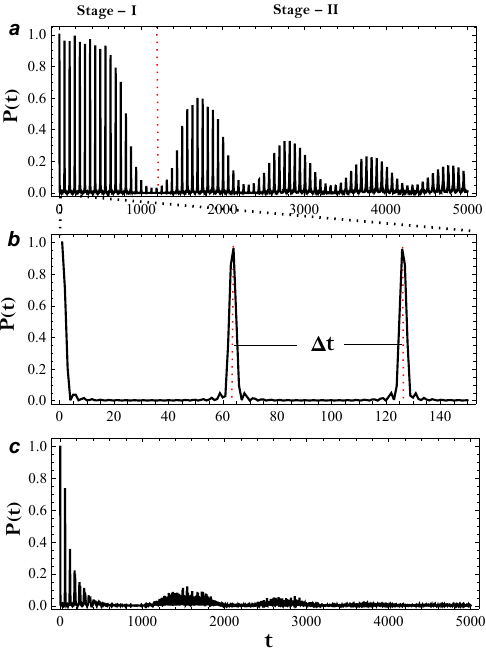}
	\end{center}
	\caption{\label{P(t)} {\bf Fidelity dynamics and its coherence revivals.} {\bf a}  The fidelity exhibits pronounced periodic revivals at early times (Stage I). At later times (Stage II), the envelope of the fidelity decays slowly. {\bf b} is an enlarged version of a part of {\bf a}, where $\Delta t$ is the revival period. {\bf c} Fidelity dynamics of an initial state constructed by independent phase randomization of each occupied single-particle orbital and spin component. The particle number is $N=40$, with interactions $c_0=1.0$ and $c_2=0.1$.}
\end{figure}

As shown in Fig.\ref{P(t)}{\bf a}, the fidelity exhibits two distinct dynamical regimes characterized by different revival envelopes. During Stage I, the system displays pronounced and nearly periodic revivals with a well-defined period, coinciding with the interval in which the relative populations exhibit strong and regular oscillations (Fig.\ref{entropy}{\bf a}). In Stage II, although the revival amplitude gradually decreases, the underlying oscillation period remains clearly discernible.

To quantify the revival periodicity, we identify local maxima of the fidelity exceeding a prescribed amplitude threshold and separated by a minimum temporal spacing, thereby suppressing high-frequency fluctuations. The peak positions are refined using quadratic interpolation over neighboring time steps to obtain sub-step accuracy. The intervals between successive maxima define the revival period. In Stage-I, this procedure yields an average peak spacing of $\Delta t \approx 62.668$ time steps with a variance of $0.039$. For a simulation time step $dt=0.1$, this corresponds to a physical revival period $T_{\mathrm{rev}} \approx 6.2668$. Repeating the analysis in Stage-II, focusing on the clearly identifiable large-amplitude revivals, gives an average spacing $\Delta t \approx 62.6924$ with a variance of $0.037$, corresponding to $T_{\mathrm{rev}} \approx 6.26924$. The close agreement between these values demonstrates that the dominant dynamical timescale remains essentially unchanged throughout the evolution. The gradual decay of the envelope therefore reflects amplitude modulation and dephasing rather than a drift of the intrinsic frequency scale.

We emphasize that the persistence of fidelity revivals in Stage II does not imply a simultaneous recovery of all observables. The fidelity is a global, phase-sensitive probe of coherent interference in the full Hilbert space, whereas the relative spin populations are coarse-grained quantities defined by projection onto specific sectors. Partial rephasing within a dynamically selected subset of modes can therefore produce pronounced fidelity peaks even when coarse observables exhibit irregular behavior. The distinct oscillatory patterns thus reflect the separation between global coherence and observable relaxation.

Without loss of generality, we construct an initial state by independently randomizing the phases associated with each orbital and spin component, while keeping the occupation configuration and amplitude distribution unchanged. Specifically, each spinor amplitude in orbital $n$ and spin component $m$ is multiplied by an independent phase factor $e^{i\theta_{n,m}}$, where $\theta_{n,m}$ are independently drawn from a uniform distribution over $[-\pi,\pi]$. This procedure preserves the single-particle occupations but removes coherent phase correlations across different orbitals and spin sectors. The resulting fidelity dynamics is shown in Fig.\ref{P(t)}{\bf c}. The fidelity exhibits rapidly damped oscillations and fast decay toward a stationary regime. Although weak residual modulations remain visible, their amplitudes are strongly suppressed and do not display long-lived coherent revivals. The residual weak modulation originates from the near-equidistant single-particle spectrum of the harmonic confinement and does not rely on the initial-state coherence.

\begin{figure}
	\begin{center}
		\includegraphics[width=0.9\linewidth]{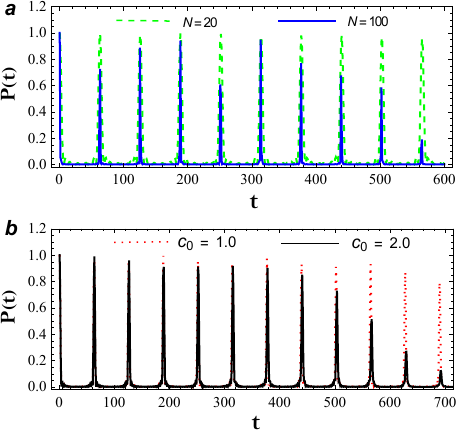}
	\end{center}
	\caption{\label{P-NC} {\bf Time evolution of the fidelity $P(t)$ for different system parameters in Stage-I.} {\bf a} Comparison of systems with particle numbers $N=20$ (green dashed line) and $N=100$ (blue line). The interactions are set to $c_0=1.0$ and $c_2=0.1$. {\bf b} Comparison of different interaction strengths $c_0=1.0$ (red dotted line) and $c_0=2.0$ (black line), for a fixed particle number $N=40$ and non-mixing interaction strength $c_2=0.1$.}
\end{figure}

Fig.\ref{P-NC} further demonstrates the robustness of the dominant dynamical timescale in Stage I. In Fig.\ref{P-NC}{\bf a}, systems with different particle numbers exhibit revival peaks occurring at essentially the same temporal positions, indicating that the intrinsic oscillation period is insensitive to system size within the explored range. However, the larger system ($N=100$) displays visibly reduced peak amplitudes, reflecting enhanced dephasing and stronger mixing among dynamical components as the Hilbert-space dimension increases. A similar trend is observed in Fig.~\ref{P-NC}{\bf b}. Increasing the interaction strength from $c_0=1.0$ to $c_0=2.0$ suppresses the revival amplitude while leaving the oscillation period nearly unchanged. This behavior suggests that stronger interactions primarily enhance phase dispersion among contributing modes, reducing the coherence of the revivals, whereas the dominant frequency scale underlying the dynamics remains largely stable.

 \begin{figure}
	\begin{center}
		\includegraphics[width=0.9\linewidth]{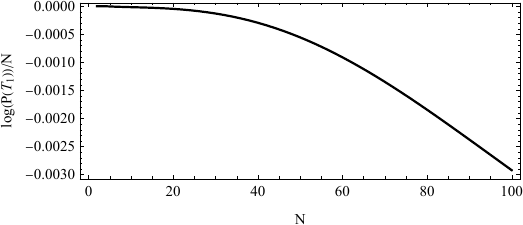}
	\end{center}
	\caption{\label{density} {\bf The fidelity density of the first revival peak.} For each particle number $N$, we extract the first prominent revival maximum $P(T_1)$ from the return probability and define the fidelity density as $\log P(T_1)/N$. The weak dependence on $N$ indicates that the revival weight does not exhibit strong exponential suppression within the accessible system sizes.}
\end{figure}

To quantify how the revival coherence scales with system size, we define a fidelity density based on the first revival peak. Specifically, we extract the first prominent maximum $P(T_1)$ of the return probability and compute the fidelity density $\log P(T_1)/N$, where $N$ is the particle number. 
As shown in Fig.\ref{density}, the fidelity density exhibits only a weak variation as $N$ increases. In generic thermalizing many-body systems, revival amplitudes typically decrease rapidly with system size, often consistent with exponential suppression. Such behavior would manifest as a pronounced downward trend in $\log P(T_1)/N$. Instead, within the accessible system sizes, we observe no strong systematic decay of this fidelity density.

 \begin{figure}
	\begin{center}
		\includegraphics[width=0.9\linewidth]{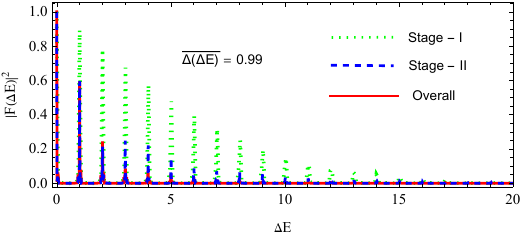}
	\end{center}
	\caption{\label{Fourier} {\bf Fourier power spectra $|F(\Delta E)|^2$ of the fidelity $P(t)$.} The Fourier spectrum of $P(t)$ is calculated for three distinct time intervals. The three spectra almost perfectly overlap, displaying a series of evenly spaced peaks, although the spectral amplitude gradually decreases. Each spectrum is normalized by its maximum value. The particle number is $N=40$ and the interactions are $c_0=1.0$ and $c_2=0.1$.}
\end{figure}

To gain further insight into the dynamical structure underlying the revivals, we analyze the Fourier spectra of the fidelity. For a selected time window $[t_1, t_2]$, the discrete Fourier transform (DFT) of the fidelity is defined as
\begin{equation}
	F(\Delta E) = \sum_{t=t_1}^{t_2} P(t)\, e^{-i \Delta E\, t\, }.
\end{equation}

The corresponding power spectrum $|F(\Delta E)|^2$ characterizes the dominant frequency components contributing to the oscillatory behavior of $P(t)$. To compare different dynamical regimes, we analyze three representative time intervals: Stage I ($t \in [1, 1000]$), Stage II ($t \in [1000, 5000]$), and the overall evolution ($t \in [1, 5000]$). As shown in Fig.\ref{Fourier}, the spectra obtained from all three intervals nearly overlap and exhibit a series of sharp, discrete peaks located at fixed quasienergy differences. The dominant peak corresponds to an effective characteristic scale $\Delta(\Delta E) \approx 0.99$, which sets the primary revival timescale of $P(t)$. This frequency should not be interpreted as a single instantaneous level spacing. Because the Hamiltonian is time dependent within the self-consistent mean-field dynamics, the relevant phase accumulation is governed by time-integrated energy differences. The observed peak therefore represents an effective interference scale arising from the collective phase evolution of multiple dynamically participating modes. In this sense, the fidelity spectrum encodes the dominant time-averaged phase difference rather than a static microscopic gap. The persistence of these discrete spectral peaks across different time windows demonstrates that the underlying frequency organization remains stable, even though the revival amplitude gradually decreases due to dephasing. Together, these features indicate the presence of a quasi-regular spectral structure embedded within the many-body spectrum that governs the long-lived oscillatory dynamics.

Building on the quasienergy analysis above, we project the initial wavefunction onto the instantaneous eigenbasis of the effective Hamiltonian $H_{\mathrm{eff}}$ at representative times. The effective Hamiltonian is defined as
\begin{equation}
	H_{\mathrm{eff}}(t) = H_{\mathrm{kin}} + H_{\mathrm{trap}} + H_{\mathrm{int}}[\rho(t)].
\end{equation}
Here, the kinetic and trapping terms are single-particle operators, while the interaction contribution depends explicitly on the instantaneous density matrix $\rho(t)$. Diagonalizing $H_{\mathrm{eff}}(t)$ in the same discretized spatial–spin basis yields the instantaneous eigenenergies $ \varepsilon_n(t)$ and eigenstates $|\phi_n\rangle(t)$,

\begin{equation}
	H_{\mathrm{eff}}(t)\,|\phi_n\rangle = \varepsilon_n(t)\,|\phi_n(t)\rangle.
\end{equation}
Although the Hamiltonian is time-dependent, this instantaneous eigenbasis provides a convenient representation for diagnosing how the evolving state distributes over the self-consistent spectrum at selected times. This construction introduces no additional approximation beyond the TDHF equation used in the real-time evolution.

The projection weights are calculated as
\begin{equation}
	w_n = |\langle \phi_n | \psi(0) \rangle|^2,
\end{equation}
which quantify the overlap of the initial state with the instantaneous eigenmodes. These weights serve as a spectral diagnostic of the interference structure underlying the fidelity dynamics.

 \begin{figure}
	\begin{center}
		\includegraphics[width=1\linewidth]{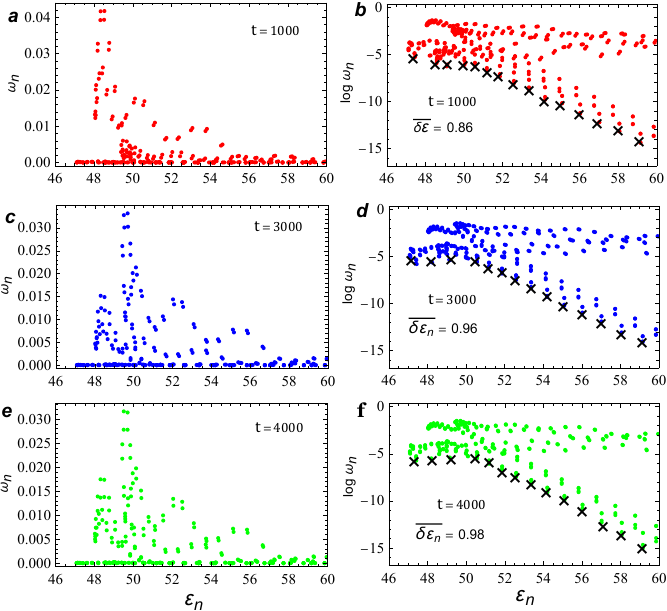}
	\end{center}
	\caption{\label{ladder} {\bf Spectral decomposition of the instantaneous effective Hamiltonian at different evolution times.}   ($\mathbf{a_1}$–$\mathbf{a_3}$) Projection weights $w_n = |\langle \phi_n | \psi(0) \rangle|^2$ as a function of eigenenergy $\varepsilon_n$, evaluated at evolution times $t = 1000$, $3000$, and $4000$.  A subset of quasi-equispaced modes (marked by ``×'') forms a structured ladder embedded in the continuum. The particle number is $N=40$ and the interactions are $c_0=1.0$ and $c_2=0.1$.}
\end{figure}

As shown in Fig.\ref{ladder}$\mathbf{a}$, $\mathbf{c}$, and $\mathbf{e}$, the projection weights are broadly distributed across many instantaneous eigenmodes, indicating that the initial state is not dominated by a small set of eigenstates with large overlaps. The overall weight profile evolves gradually over time, reflecting progressive dephasing within the many-body spectrum. The logarithmic plots in Fig.\ref{ladder}$\mathbf{b}$, $\mathbf{d}$, and $\mathbf{f}$, reveal a subset of modes (marked by ``×''). Notably, the positions of these marked modes along the energy axis and their mutual spacing exhibit some drift between $t=1000$, $t=3000$, and $t=4000$, corresponding to different dynamical stages. Despite this drift, the quasi-equispaced structure appears to remain stable. To further investigate this, we increased the time resolution by calculating the instantaneous spectrum every 100 time steps. By averaging the energy differences across these different time points, we find that the resulting average value $\overline{\delta \varepsilon_t} \simeq 0.97$ is comparable to the energy difference obtained from the Fourier spectrum. This suggests that the energy spacing within the ladder of instantaneous eigenstates cannot be determined from any single time snapshot. Rather, it arises from a time-averaged structure that reflects the collective phase evolution over the entire dynamical trajectory. Although these modes do not individually carry the largest projection weights, their near-uniform spacing enables coherent phase interference over long timescales. Importantly, the dynamical contribution of a subset of modes is not determined solely by the magnitude of their individual overlaps, but by the coherence of their energy differences. Modes with larger projection weights but irregular level spacings generate a broad distribution of frequencies and predominantly contribute to dephasing. In contrast, a quasi-regular subset with moderate overlaps can dominate the revival dynamics when many pairwise energy differences cluster around a common value, leading to constructive phase interference. The proximity between this quasi-regular spacing and the effective frequency scale extracted from the fidelity spectrum indicates that the observed revivals originate from the collective interference of these weakly hybridized modes. The consistency between the Fourier analysis and the instantaneous spectral structure supports the interpretation that the long-lived oscillations arise from a quasi-regular spectral structure embedded within the many-body continuum. While this spectral organization shares phenomenological similarities with quantum many-body scars, it differs from an idealized eigenstate-dominated mechanism and instead reflects collective interference among a sparse set of dynamically selected modes.

 \section{Conclusion}
 
 In summary, we have investigated the nonequilibrium dynamics of a harmonically trapped spin-$3/2$ Fermi gas within the time-dependent Hartree–Fock equation. Starting from a spin-imbalanced initial state, we analyzed the evolution of the Shannon entropy and the fidelity to characterize weak ergodicity breaking. The entropy exhibits long-lived oscillations and gradually moves toward thermal behavior at long times, indicating delayed equilibration rather than permanent confinement. The fidelity shows robust revivals with a nearly constant period that remains stable across different system sizes and interaction strengths, while the gradual reduction of revival amplitude reflects progressive dephasing. Spectral analysis reveals that this stable dynamical timescale is associated with a quasi-regular, scar-like subset of modes embedded in the dense many-body spectrum. Although these modes do not dominate the overlap distribution, their approximate energy regularity enables coherent phase interference that underlies the observed revivals. Overall, our results show that long-lived nonthermal oscillations in this continuum system originate from a scar-like spectral organization within the many-body background. These findings highlight how such scar-like spectral structure can shape nonequilibrium dynamics beyond lattice models and without exact constraints.

\begin{acknowledgments}
	This work was supported by the National Natural Science Foundation of China (Grant Nos. 11574028 and 11874083).
\end{acknowledgments}

\end{document}